# Cognition and Reality

F. Tito Arecchi

Emeritus of Physics-Università di Firenze

and

 INO-CNR –Firenze

Abstract

We discuss the two moments of human cognition, namely, *apprehension* (A),whereby a coherent perception emerges from the recruitment of neuronal groups, and *judgment*(B),that entails the comparison of two apprehensions acquired at different times, coded in a suitable language and retrieved by memory. (B) entails *self-consciousness,* in so far as the agent who expresses the judgment must be aware that the two apprehensions are submitted to his/her own scrutiny and that it is his/her task to extract a mutual relation. Since (B) lasts around 3 seconds, the semantic value of the pieces under comparison must be decided within that time. This implies a fast search of the memory contents.
As a fact, exploring human subjects with sequences of simple words, we find evidence of a limited time window , corresponding to the memory retrieval of a linguistic item in order to match it with the next one in a text flow (be it literary, or musical,or figurative).

While *apprehension* is globally explained as a Bayes inference, *judgment t*results from an inverse Bayes inference. As a consequence, two hermeneutics emerge (called respectively *circle* and *coil*). The first one acts in a pre-assigned space of features. The second one provides the discovery of novel features, thus unveiling previously unknown aspects and  hence representing the road to reality.

   *Outline*

   *1-Perception, judgment and self-consciousness*

   *2-The brain operations-Role of homoclinic chaos*

   *3- Perception as a Bayes inference*

   *4- Linguistic operations as inverse Bayes*

  *5-Two different hermeneutics, that is, interpretations of cognitive data*

   *6- Conclusions- Two aspects of linguistic creativity*

   *Bibliography*

## 1-Perception, judgment and self-consciousness

Figs 1 and 2 introduce the difference between *A-apprehension* or *perception* that rules the motor reactions of any brainy animal, and *B-language*, only humans, and that provides judgments.

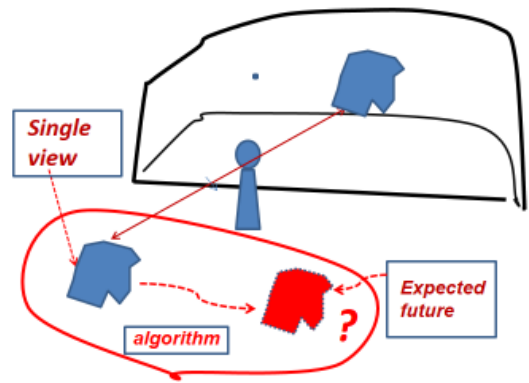

*Fig.1-Plato said that we see the shadows of things, like a prisoner constrained to view the end of a cave and forbidden to turn and see the outside world. This occurs indeed in perceptual tasks, where the sensorial stimuli are interpreted by "algorithms" and generate (within1 sec) a motor reaction. The procedure is common to all brainy animals.*

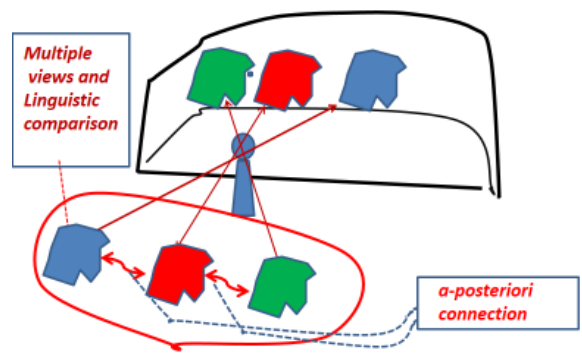

*Fig.2- In linguistic operations, humans code a perception in a linguistic code andretrieve it by short term memory (around 3 sec)comparing it to a successive coded perception. From the comparisonit emerges a connection that increases the details of the observed thing.*

Figs 3 and 4 show why the scientific program is a linguistic one and what is the reason of its success.

With this in mind, we explore whether and how Cognition unveils Reality ..

Following the philosophy of cognition of Bernard Lonergan [Lonergan], I discuss two distinct moments of human cognition, namely, *apprehension* (*A*) whereby a coherent perception emerges from the recruitment of neuronal groups, and *judgment (B)* whereby memory recalls previous (A) units coded in a suitable language; these units are compared and from comparison it follows the formulation of a judgment.

The first moment, *(A),* has a duration around 1 sec; its associated neuronal correlate consists of the synchronization of the EEG (electro-encephalo-graphic ) signals in the so-called gamma band (frequencies between 40 and 60 Hz) coming from distant cortical areas .It can be described as an interpretation of the sensorial stimuli on the basis of available algorithms, through a Bayes inference.

Precisely, calling *h* (h= hypothesis) the interpretative hypotheses in presence of a sensorial stimulus *d* (d=datum), the Bayes inference selects the most plausible hypothesis *h\**,that determines the motor reaction, exploiting a memorized algorithm *P(d|h),* that represents the conditional probability that a datum *d* be the consequence of an hypothesis *h.*

The *P(d|h)* have been learned during our past; they represent the equipment whereby a cognitive agent faces the world. By equipping a robot with a convenient set of *P(d|h),* we expect a sensible behavior.

The second moment, *(B)*,entails a comparison between two apprehensions (A) acquired at different times, coded in a given language and recalled by the memory. If, in analogy with (A), we call *d* the code of the second apprehension and *h\** the code of the first one, now- at variance with (A)- *h\** is already given; instead, the relation *P(d|h)* which connects them must be retrieved; it represents the *conformity* between *d* and *h\*,* that is, the best interpretation of *d* in the light of *h\*.*

Thus, in linguistic operations, we compare two successive pieces of the text and extract the conformity of the second one on the basis of the first one. This is very different from (A), where there is no problem of conformity but of plausibility of *h\** in view of a motor reaction.

Let us make two examples: a rabbit perceives a rustle behind a hedge and it runs away, without investigating whether it was a fox or just a blow of wind.

On the contrary, to catch the meaning of the 4-th verse of a poem, we must recover the 3-d verse of that same poem, since we do not have a-priori algorithms to provide a satisfactory answer.

Once the judgment, that is, the *P(d|h)* binding the codes of the two linguistic pieces in the best way, has been built, it becomes a memorized resource to which to recur whenever that text is presented again. It has acquired the status of the pre-learned algorithms that rule (A)

However-at variance with mechanized resources- whenever we re-read the same poem, we can grasp new meanings that enrich the previous judgment *P(d|h)*. As in any exposure to a text (literary, musical, figurative) a re-reading increases our understanding.

(B) requires about 3 seconds and entails *self-consciousness,* as the agent who expresses the judgment must be aware that the two successive apprehensions are both under his/her scrutiny and it is up to him/her to extract the mutual relation.[Arecchi, 2007, Doyia et al. ]

At variance with (A), (B) does not presuppose an algorithm, but rather it builds a new one through an *inverse Bayes procedure* [Arecchi2007]. This construction of a new algorithm is a sign of *creativity* and *decisional freedom*

Here the question emerges: can we provide a computing machine with the (B) capacity, so that it can emulate a human cognitive agent, as expected in the Turing test? .The answer is NOT, because (B) entails non-algorithmic jumps, insofar as the inverse Bayes procedure generates an *ad hoc* algorithm, not available previously.

The scientific endeavor can not be carried on by an AI (artificial intelligence ) device, since it entails a linguistic step, as shown in Fig,3. Fig.4 explains why Galileo's program provides certainties, rather than probabilistic expectations.

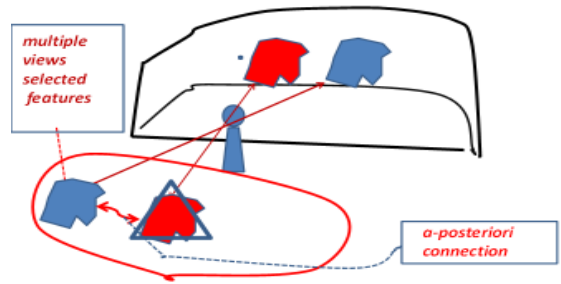

*Fig.3-The scientific program is a linguistic task. Galileo's approach consists of extracting mathematical features; it implies a linguistic operation, according to Fig.2*

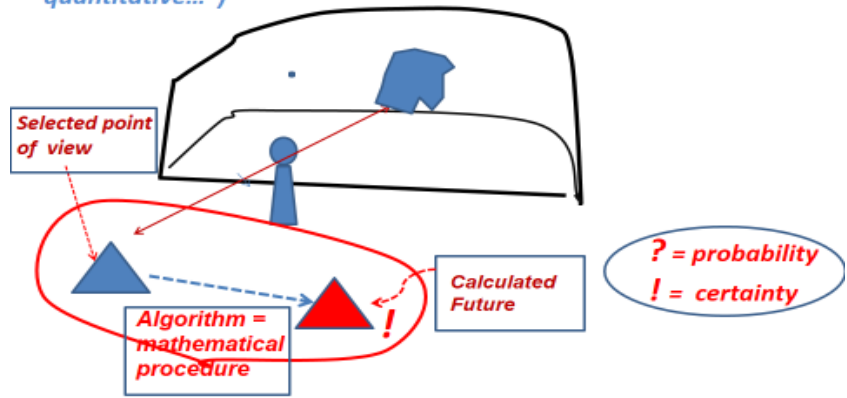

*Fig.4-Once the object under investigation is reduced to a collection of mathematical features, one applies millennia of mathematical wisdom and predicts the future behavior .*

## 2-The brain operations - Role of homoclinic chaos

Let us introduce "deterministic chaos".[Arecchi,2004 b] Since Poincaré (1890) we know that a dynamical system is extremely sensitive to the initial conditions. That yields the so called "butterfly effect" whereby a tiny shift in the initial conditions yields a large difference in course of time . Precisely, a difference of initial conditions induces a divergence of the dynamical trajectories in course of time. In the case of a meteo dynamical model, accounting for most atmospheric features (wind, pressure, humidity,etc) but neglecting the motion of a butterfly wing could lead to a wrong prediction(from sunny to rainy).Loss of the initial information occurs over a time $\tau$ whose inverse is called **$K$**(Kolmogorov entropy). In the meteo model such a $\tau$ may be days, in a dynamical model of the solar system it takes millions of years

We call *geometric chaos* the above trajectory divergence.

Another type of chaos, that we call *temporal chaos,* consists of regular closed orbits that however repeat at irregular times (Fig.5).

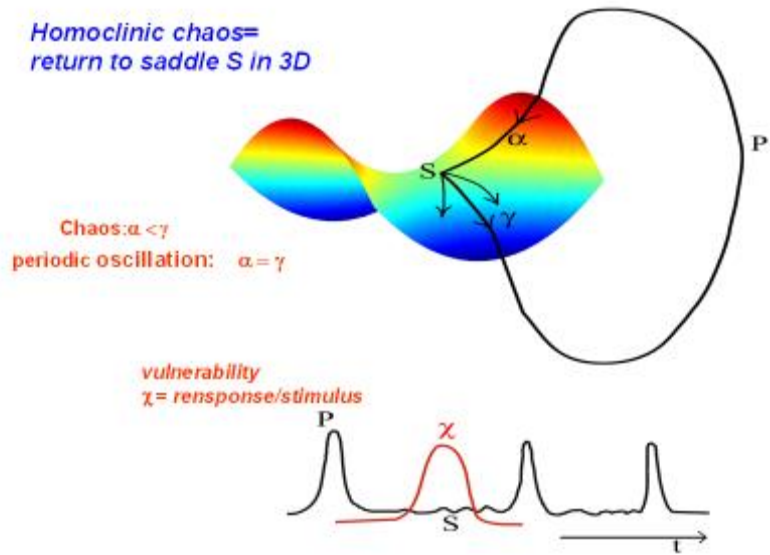

*Fig.5-Homoclinic chaos; the dynamic trajectory is a closed orbit starting from S (saddle point) and returning to it. Projecting on a single direction, we observe spikes P repeating in time. The time separation between two spike occurrences depends on the relations between α and γ, thus it can be controlled by a voltage applied to S, as the signal $\chi$ .*

A single neuron in the brain undergoes temporal chaos and its electrical output consists of a train of spikes (each one high 100mV and lasting 3 ms). The minimal inter-spike separation is 3 ms; the average separation is 25 ms in the so-.called γ band of the EEG(electro-encephalo-gram).

A neuron communicates with other neurons in two ways [Arecchi,2004 a, Singer, Womelsdorf. & Fries]:

-either directly , by coupling its spike train to another neuron via an electric line called *axon,*

-or indirectly, by building with nearby neurons a local potential (detectable as an EEG signal) and providing a signal χ to a distant neuron, that consequently re-adjusts its firing rate.

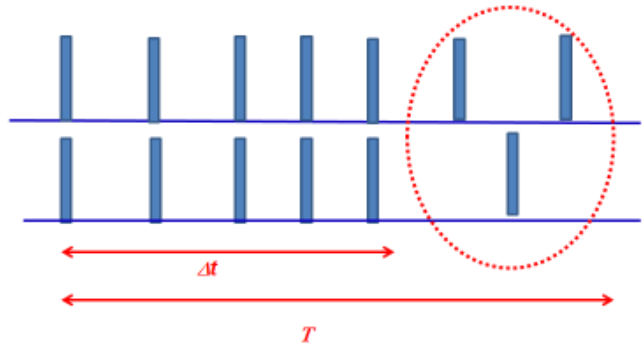

*Fig.6- .Direct coupling of two neurons by synchronized spike trains; synchronization missed after Δt for an extra-spike in the upper train.*

Fig.6 shows the direct synchronization of two trains over a time Δt. The neurons involved in the coupling are confined in a thin layer of the brain (thickness 2 mm) called the *cortex*. Groups of nearby neurons contribute to a common task forming a specialized area that builds global interactions with other areas (Fig.7).

The areas are visualized via the amount of oxygenated blood required by a working region and visualized by *f- MRI*(functional magnetic resonance imaging).

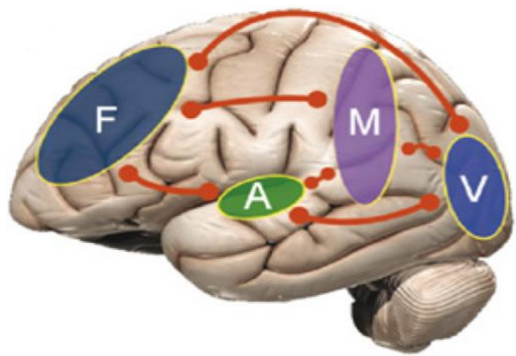

**multisensory interactions**
will combine into a unified pattern involving frontal cortex, temporo-parietal regions as well as unimodal cortices: **A** = auditory cortex; **V** = visual cortex; **M**= higher-order multisensory regions; **F** = prefrontal cortex

*Fig.7- Topology of specialized cortical areas , each one being active as a large collection of synchronized neurons; mutual communication occurs via EEG signals*

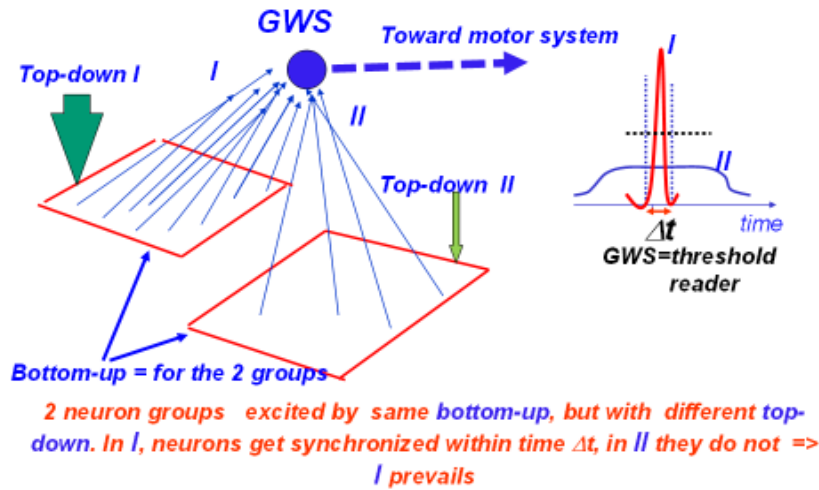

*Fig. 8- . Competition of two cortical areas with different degrees of synchronization*

Fig.8 visualizes the competition between two neuron groups **I** and **II** fed by the same sensorial (**bottom-up**) stimulus, but perturbed (**top-down**) by different interpretational stimuli provided by the long term memory. **I** wins, as the corresponding top-down stimulus succeeds in synchronizing the neuron pulses of this group better than in group **II**. This means that, over a time interval $\Delta t$, neurons of **I** sum up coherently their signals, whereas neurons of **II** are not co-ordinated, hence yielding a smaller sum. As a consequence a **reader GWS (= global workspace**, name given to the cortical area where signals from different areas converge; it is located in area F of Fig.7) reads within $\Delta t$ a sum signal overcoming a suitable threshold and hence eliciting a motor response [Dehaene].
Thus, the winning interpretation driving the motor system is that provided by **I**.
What represented in Fig. 8 models the mechanism *(A)* common to any animal with a brain.

## *4-Perception as a Bayes inference*

Neurosciences hypothesize a collective agreement of crowds of cortical neurons through the mutual synchronization of trains of electrical pulses (spikes) emitted individually by each neuron [ Singer et al., Dehaeneet al.] .The neuroscientific approach is summarized in Fig.8.
However, a global description of the above process can be carried on in probabilistic terms, without recurring to the details of the process.
In 1763, Thomas Bayes, looking for a reliable strategy to win games, elaborated the following probabilistic argument[Bayes]. Let us formulate a manifold of hypotheses *h* about the initial situation of a system, attributing to each hypothesis a degree of confidence expressed by an a priori probability
$$P(h).$$
Any hypothesis, introduced as input into a *model* of evolution, generates data. Let us assume that we know the model and, hence, can evaluate the probability of the *data conditioned* by a specific hypothesis *h*; we write it as
$$P(data|h).$$

The model is like an instruction to a computer, thus we call it *algorithm*; it generates different data for different *h*. If then we perform a measurement and evaluate the probability

*P(data)*

of the data, we must conclude that there is an *h* more plausible than the other ones, precisely the one that maximizes the probability conditioned by the data

*P(h|data)*,

that we call the a posteriori probability of *h* and denote as *h\**.

This procedure is encapsulated in the formula, or theorem, of Bayes, that is

*P(h\*)=P(h|data) = P(h) [P(data|h)/P(data)]*

To summarize, the a posteriori probability of *h*, conditioned by the observed data, is given by the product of the a priori probability of *h*, times the probability *P(data|h)* of the data conditioned by a given *h*, that we call the *model,* and divided by the probability *P(data),* based on a previous class of trials.(Fig.9)

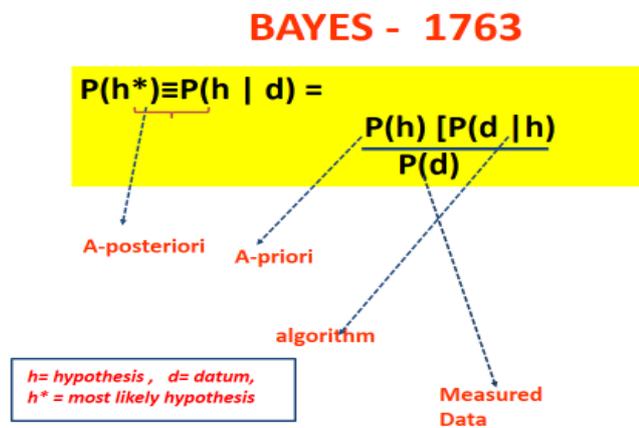

*Fig.9 Bayes inference*

Fig.10 summarizes the whole perception procedure, that is initiated by an external stimulus and concluded by a motor reaction.

. Successive applications of the theorem yield an increasing plausibility of *h\**; it is like climbing a mountain of probabilities along its maximum slope, up to the peak. After each measurement of the data and consequent evaluation of the a posteriori *h\**, we reformulate a large number of new a priori *h* relative to the new situation, and so on.(Fig.11). . Notice that Darwinian evolution by *mutation* and successive *selection* of the best fit mutant is a sequential implementation of Bayes theorem.

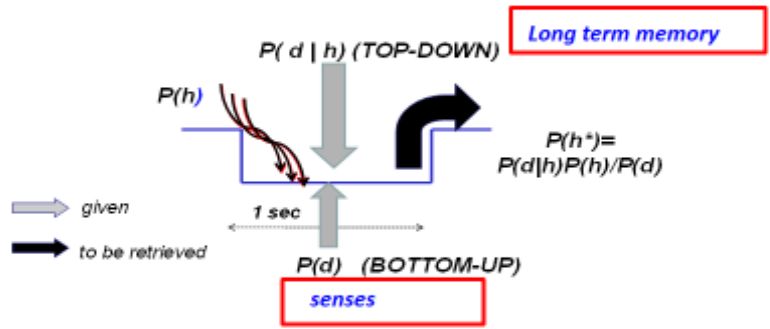

*Fig. 10-Starting with a large number of presumed hypotheses h, the occurrence of the data selects the h\* that satisfies the above relation and drives a suitable reaction.*

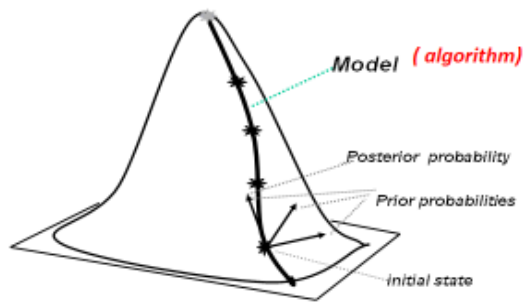

*Fig.11.Recursive application of Bayes is equivalent to climbing a probability mountain, guided by the Model ,that is, the conditional probability that an hypothesis generates a datum.
This strategy is common e.g. to Darwin evolution and to Sherlock Holmes criminal investigation; since the algorithm is unique, it can be automatized in a computer program (expert system)*

## 4- Linguistic operations as inverse Bayes

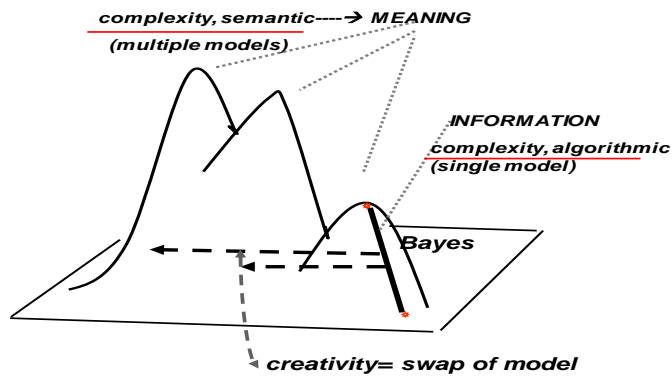

**Climbing up a single peak is a non-semiotic procedure**
**ON THE CONTRARY**
**Jumping to other peaks is a creativity act, implying a holistic comprehension of the surrounding world (semiosis)**

*Fig.12. Comparison of two different complexities, namely, i) the algorithmic C., corresponding to the bit length of the program that enables the expert system to a recursive Bayes; and ii) semantic C., corresponding to the occurrence of different models*

In Fig.11 the recursive application of Bayes using the same algorithm- or model- is visualized as climbing a probability mountain. The bit length of the algorithm is the *Algorithmic Complexity* of the cognitive task.

However, in everyday life we experience jumps toward different algorithms, that means going to climb different mountains (Fig.12). The associated multiplicity of choices corresponds to attributing different meanings to the input data; the number of alternative choices will be called *Semantic Complexity*.

This swap of the model is a creative jump proper of language operations.

It is the root of Goedel-1931 incompleteness theorem and Turing-1936 halting problem for a computer, as discussed in a previous paper [Arecchi2012].

Altogether different from *(A)* is the situation for *(B),* that- implying the comparison between different apprehensions coded in the same language (literary, musical, figurative, etc.)- represents an activity exclusively human.
In fact, the second moment *(B)* entails the comparison of two apprehensions acquired at different times, coded in the same language and recalled by the memory.
*(B)* lasts around 3 sec; it requires *self-consciousness,* since the agent who performs the comparison must be aware that the two non simultaneous apprehensions are submitted to his/her scrutiny in order to extract a mutual relation.

At variance with *(A), (B)* does not presuppose an algorithm but it rather builds a new one through an *inverse Bayes procedure* introduced by Arecchi [Arecchi,2007]. This construction of a new algorithm is the source of *creativity* and *decisional freedom.*
Language indeed permits an infinite use of finite resources [Humboldt].
It is the missing step in Turing's claim that human intelligence can be simulated by a machine [Turing].
The first scientist who explored the cognitive relevance of the 3sec interval has been Ernst Pöppel [Pöppel ].
This new temporal segment has been little explored so far. All the so-called *"neural correlates of consciousness"(NCC)* are in fact electrical *(EEG)* or functional magnetic resonance *(fMRI)* tests of a neuronal recruitment stimulating a motor response through a GWS (see Fig.8); therefore they refer to (A). In such a case, rather than *consciousness* , one should call it *perceptual awareness,* that we have in common with brainy animals.

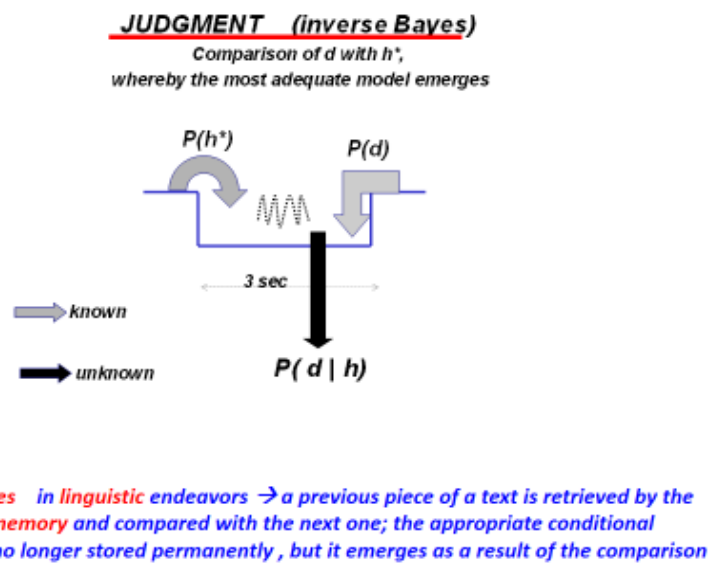

*Fig. 13-The inverse Bayes procedure that occurs in linguistic endeavors, whereby a previous piece of a text is retrieved by the short term memory and compared with the next one: the appropriate conditional probability is no longer stored permanently but it emerges as a result of the comparison (judgmentand consequent decision).*

Fig.13 shows how an inverse Bayes procedure provides the best comparisons of two successive pieces of a linguistic text, thus generating a judgment.
While in perception we compare sensorial stimuli with memories of past experiences, in judgment we compare a piece of a text coded in a specific language (literary, musical, figurative) with the preceding piece, recalled via the short term memory. Thus we do not refer to an event of our past life, but we compare two successive pieces of the same text.
Such an operation requires that:
i) The cognitive agent be aware that he/she is the same examiner of the two pieces under scrutiny;
ii) The interpretation of the second piece based upon the previous one implies to have selected the most appropriate meanings of the previous piece in order to grant the best conformity (from a technical point of view, this conformity is what in the philosophy of cognition
of Thomas Aquinas was defined as truth= ***adaequatio intellectus et rei*** *(loosely translated as* **:** *conformity between the intellectual expectation and the object under*

*scrutiny)*

In Fig. 10 we have generically denoted as *top-down* the bunch of inner resources ( emotions, attention) that, upon the arrival of a bottom-up stimulus, are responsible for selecting the model *P(d/h)* that infers the most plausible interpretation *h\** driving the motor response. The *focal attention* mechanisms can be explored through the so-called **NCC** *(*Neural Correlates of Consciousness) [Koch] related to EEG measurements that point the cortical areas where there is intense electrical activity producing spikes, or to f-MRI (functional magnetic resonance imaging) that shows the cortical areas with large activity which need the influx of oxygenated blood.
Here one should avoid a current confusion. The fact that a stimulus elicits some emotion has NOTHING to do with the judgment that settles a linguistic comparison. As a fact, **NCC** does not reveal self-consciousness, but just the awareness of an external stimulus to which one must react. Such awareness is common to animals, indeed many tests of NCC are done on laboratory animals.

It is then erroneous to state that a word isolated from its context has an aesthetical quality because of its musical or evocative power. In the same way, it is erroneous to attribute an autonomous value to a single spot of color in a painting independently from the comparison with the neighboring areas. All those "excitations" observed by fMRI refer to emotions related to apprehension and are inadequate to shed light on the judgment process.

The different semantic values that a word can take are associated with different emotions stored in the memory with different codes (that is, spike trains). Among all the different values, the cognitive operation "judgment" selects that one that provides the maximum synchronization with the successive piece.

Thus emotions are necessary but not sufficient to establish a judgment. On the other hand, emotions are necessary and sufficient to establish the apprehension as they represent the algorithms of the direct Bayes inference. This entails a competition in GWS (Fig.8) ,where the winner is the most plausible one; whereas in the judgment- once evoked the panoply of meanings to be attributed to the previous piece- these meanings do not compete in a threshold process, but they must be compared with the code of the next word in order to select the best interpretation.

Recent new terms starting with ***neuro-***( as e.g. ***neuro-ethics, neuro-aesthetics, neuro-economy, neuro-theology***) smuggle as shear emotional reactions decisions that instead are based on judgments. The papers using those terms overlook the deep difference between apprehensions and judgments.

A very successful neurological research line deals with ***mirror neurons***, that is, neurons that activate in subjects (humans or higher animals) observing another subject performing a specific action, and hence stimulate mimetic reactions [Rizzolatti]. Here too, we are in presence of mechanisms (empathy) limited to the emotional sphere, that is, very useful for formulating an Apprehension, but not a Judgment.

*5-Two different hermeneutics, that is, interpretations of cognitive data*

Fig.14 shows how a cognitive agent *A* reads an object *B*. The *CIRCLE* refers to a Bayes cognition, whereby an algorithm is taken as necessary and sufficient to generate knowledge of B.

Whenever *A* reconsiders *B*, he/she finds the same *B* already memorized.

On the contrary, expressing the knowledge in a language and comparing successive pieces by inverse Bayes, entails an increase of details of *B (B1,B2*, etc.) that improve the cognition of the agent *(A1, A2*, etc).

.

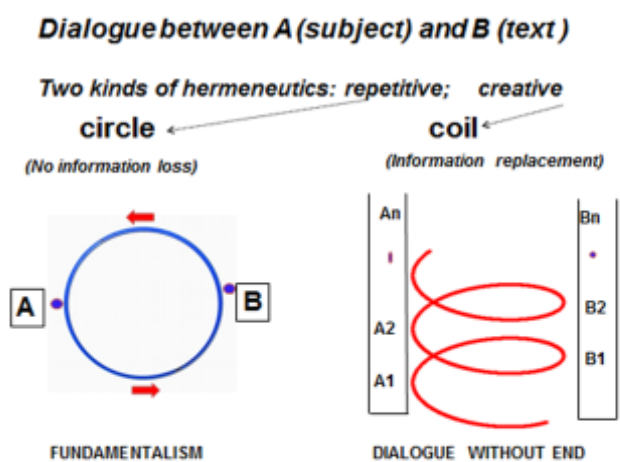

*Fig.14- Two kinds of interpretation of a text, or hermeneutics, namely, the*
*CIRCLE, whereby the interpreter A attributes a finite and fixed set of meanings to the text B, and*
*the COIL, whereby A captures some particular aspects of B and- based on that information- A*
*approaches again the text B discovering new meanings. The novel insight provided at each coil is*
*an indication of how language provides new semantic apertures,*

As for the *CIRCLE*, in information science, an **ontology** is a formal definition of the properties, and mutual relationships of the entities that exist for a particular domain of discourse. An ontology lists the variables needed for some set of computations and establishes the relationships between them. For instance, the booklet of the replacement parts of a brand of car is the ontology of that car. The fields of artificial intelligence create ontologies to limit complexity and to organize information. The ontology can then be applied to problem solving. Nothing is left out; we call this cognitive approach *" finitistic"* as no new insight is provided by repeated trials.

.
On the contrary, in any human linguistic endeavor (be it literary, or musical or figurative) *A* starts building a provisional interpretation *A1* of the text ; whenever *A* returns to *B*, he/she has already some interpretational elements to start with, and from there *A* progresses beyond , grasping new aspects *B2, B3*…and hence going to *A2* and so on *(COIL).* To carry on a COIL program, we do not need a large amount of resources; language makes an infinite use of finite resources [Humboldt].

The *COIL* hermeneutics describes also the inter-personal dialogue. If the object *B* of cognition is a human person as *A*, then the changes *B1, B2,* etc are not only due to an increased knowledge by *A*, but also to an activity of *B* who re-adjusts his/her relation with *A.*

Thus, if *B* is another human subject, then *B* undergoes similar hermeneutic updates as *A*; this is a picture of the dialogical exchange between two human beings.(persons).

*6-Conclusions- Two aspects of linguistic creativity*

We conclude by stressing two well known aspects of linguistic creativity. First, if we start a linguistic endeavor , a wealth of possible situations emerge , giving rise to ambiguous behaviors as it occurs in most products of human creativity, that –like the Etruscan Chimera- display apparently contradictory behaviors, from Ulysses to Dom Quixote (Fig.15). The onset of Chimeras is explained in Fig.16as the lack of an external referent *B*.

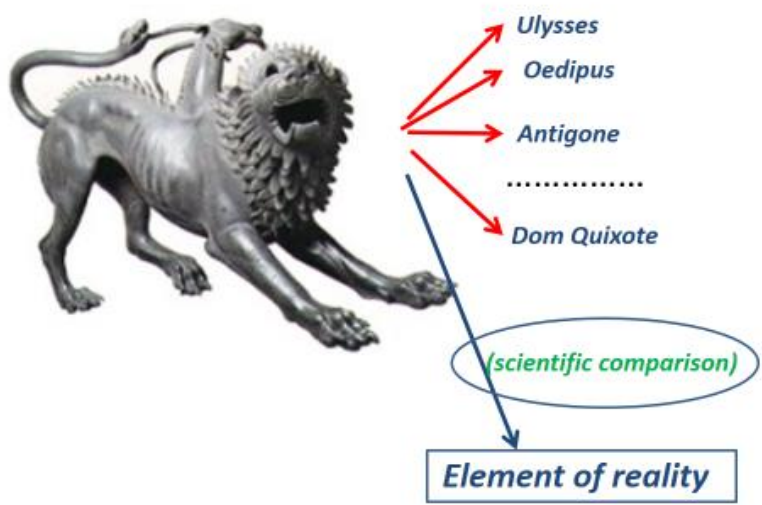

*Fig.15-A linguistic action that proceeds from a known piece toward an unknown one is like the Etruscan Chimera: it can generate mutually conflicting behaviors , as it occurs in most characters ,from Ulysses to Dom Quixote. When instead the linguistic comparison regards two observed items (as it occurs in reading the verses of a Poem, but also in scientific observations), then we really increase our personal knowledge with an element of reality.*

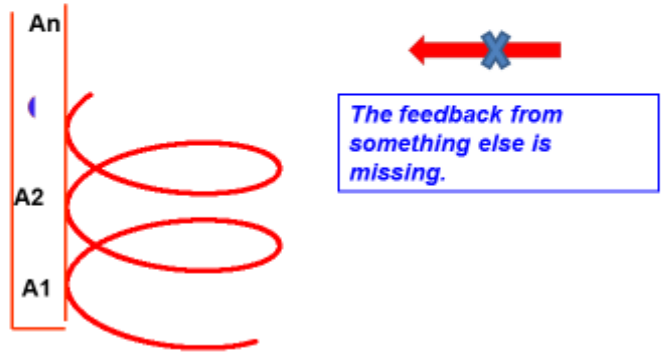

*Fig.16- How chimeras emerge in any linguistic creation*

Altogether different is the what takes place when the language is interpreting scientific observations, indeed , the repeated comparison extracts elements of reality, as hinted in the COIL hermeneutics.

Applying our hermeneutics to the scientific program, we have two possible approaches. As the size of the observed world increases from a few particles to many , within a universal scientific description associated with a fixed-algorithm (as an AI tool would operate) we witness an exponential increase of the size $C$ of the computation (that we have already called the *algorithmic complexity*) as well as a reduction of the time interval $\tau$ over which predictions are reliable, that is, an increase of the Kolmogorov entropy $K=1/\tau$. (Fig.17)

A more efficient scientific program consists of linguistic comparisons of different situations, with the help of inverse Bayes inference, applying non-algorithmic jumps as the horizontal lines of Fig. 12. Such a change of paradigm [Kuhn] leads to novel theories with low $C$ and $K$, called *effective science* [Hartmann]. A very familiar example is the formulation of Maxwell's electromagnetic equations, unifying electric, magnetic and optical phenomena.

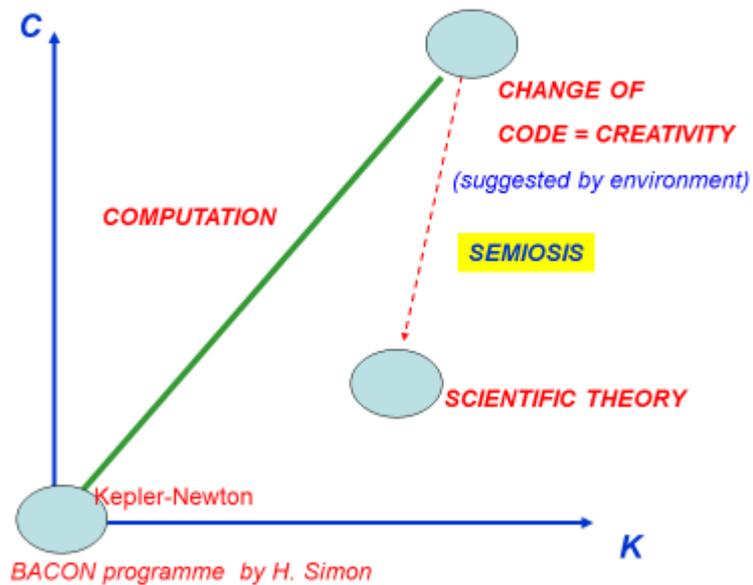

*Fig.17- Normal science vs. paradigm shift ➔effective science [Hartmann, Kuhn]-. Let **C** be the bit length of the algorithm and **K** the Kolmogorov entropy, i.e., the inverse of the time τ beyond which the initial information is lost by dynamical chaos. A simple computer program that evaluates Kepler's orbits, as BACON, has small C and **K**. As the physical system gets richer, both C and K increase and a scientific search carried on by an AI system would be affected by higher and higher C and **K**. However, a linguistic actor as a human scientist can act by a " jump of paradigm", that is, change code and introduce a new scientific theory (effective description;) with low C and K.*